\documentstyle[12pt,fleqn,twoside,epsf]{article}
\newcommand{\beq}{\begin{equation}}
\newcommand{\enq}{\end{equation}}
\newcommand{\beqa}{\begin{eqnarray}}
\newcommand{\beqast}{\begin{eqnarray*}}
\newcommand{\enqa}{\end{eqnarray}}
\newcommand{\enqast}{\end{eqnarray*}}
\newcommand{\nn}{\nonumber}
\newcommand{\lb}{\label}
\newcommand{\rf}{\ref}
\newcommand{\ct}{\cite}
\newcommand{\pa}{\partial}

\newcommand{\LD}{{\cal L}}

\newcommand{\cD}{{\cal D}}

\newcommand{\cN}{{\cal N}}

\newcommand{\al}{\alpha}
\newcommand{\be}{\beta}
\newcommand{\ga}{\gamma}
\newcommand{\de}{\delta}
\newcommand{\ep}{\epsilon}

\newcommand{\et}{\eta}

\newcommand{\la}{\lambda}
\newcommand{\rh}{\rho}
\newcommand{\si}{\sigma}

\newcommand{\ps}{\psi}

\newcommand{\De}{\Delta}

\newcommand{\La}{\Lambda}

\newcommand{\Ph}{\Phi}

\newcommand{\Ps}{\Psi}

\title{\bf Few-body aspects of non-perturbative QCD}

\author{{\bf H.G. Dosch}\\
\small Institut f\"ur Theoretische Physik der
Universit\"at Heidelberg,\\
 \small Philosophenweg 16, D69120 Heidelberg, BRD\\
\footnotesize e-mail:
h.g.dosch@thphys.uni-heidelberg.de\\[5mm]
\small Plenary Talk presented at the \\
15th International Conference on Few-Body Physics\\
\small 22-26 July 1997, Groningen, The Netherlands}
\begin{document}
\sloppy
\maketitle

\begin{abstract}
After some general remarks on non-perturbative QCD I present
shortly models which lead to a color-electric flux tube formation. The
implications of such a flux tube formation especially on high energy
scattering are discussed.
\end{abstract}

\section{Introduction}

Few body aspects of non-perturbative QCD cover a wide field , and
strictly speaking 
only meson spectroscopy can be considered to lay outside this domain,
since even baryon spectroscopy in non-perturbative QCD is at least a
three body problem because of the three valence quarks inside the the
baryon, not to speak of hadron hadron scattering, which in the simplest
case is a four body problem. I can therefore treat only a small
subsection of the topics which deserve to be considered. Since
furthermore we have no canonical analytical procedure to treat
non-perturbative QCD this selection will be necessarily a biased one.
My presentation is organized as follows:

1) Some essential features of QCD which are needed
for further reference are repeated  and 
the necessity of a non-perturbative treatment in the application to
hadron physics is pointed out.

2) A picture which is common to several
attempts to 
understand confinement -- the most intriguing feature of non-perturbative
QCD -- is presented.
It is  the picture in which a color-electric string is formed between
the quarks and (anti-) quarks. 
 Some treatments of
non-perturbative QCD which lead to such a string picture are also very
shortly exhibited.

3) A simplified version of this picture is given and  some
consequences of it are discussed, especially baryon spectroscopy and
the formation of exotic states.

4) In the last point some typical few body aspects in high
energy scattering are presented.

\section{ Some essential features of QCD}

We all tend to agree that the Lagrangian describing strong interactions
is known, namely that of quantum chromodynamics, the theory of colored
quarks and gluons.

It looks conspicuously similar to that of the
time honored theory of electrons and photons, i. e.
quantum electrodynamics. 
\beq
\LD_{QCD}(x) = i \bar \psi^{C f}(x)  \ga^\mu D_\mu^{C E} \psi^{E f}(x)
- m_f \bar \psi^{C f}(x)\psi^{C f}(x)
- \frac{1}{4} F^{\mu\nu F}(x) F_{\mu\nu}^F(x) .  
\lb{2.1}\enq

Here $\psi^{E f}(x)$ is the quark field with color $E$ and flavor $f$,
$F_{\mu\nu}^F(x)$ is the gluon field tensor defined as:
\beq F_{\mu\nu}^F:= \pa_\mu A_\nu^F - \pa_\nu A_\mu^F 
+ g_s f_{GHF} A_\mu^G A_\nu^H \lb{2.5} \enq

with the color potential $A^F_\mu(x)$ and the covariant derivative:
\beq
D_\mu^{C E} := \pa_\mu \de^{C E} -i g_s/2  \la^{C E} _F A_\mu^F
\lb{2.1a} \enq
where $\la_F$ are the Gell-Mann matrices.

I just remind you that the nonlinear term in the gluon field-tensor, which is
not present in the electro-dynamic field strength tensor, leads 
through the last term in (\rf{2.1}) to cubic and
quartic terms in the potentials $A^F_\mu(x)$ in the QCD-Lagrangian and thus
to non-linear terms in the equivalents of the Maxwell equations.

The Lagrangian \rf{2.1} has been constructed in such a way that it
is invariant if the quark and gluon fields are
transformed by  space-time dependent
gauge transformations of the group $SU(3)$. The non-linear terms are a
necessary consequence of this local gauge invariance under a non-Abelian transformation.

Analogously
to the path integral formulation of quantum mechanics, we obtain the
quantum mechanical expectation value of a functional  of quark and
gluon fields $F[\psi,A]$ as a functional integral which we write formally as:
\beq
<F[\ps,A]> = \frac{1}{\cN}\int \cD \psi \cD \bar\psi \cD A \,
F[\psi, A] e^{iS_{QCD}[\psi,\bar \psi, A]}
\lb{2.6d} \enq
where the normalization constant $\cN$ is given by the integral without the function
$F$:
\beq
\cN=\int \cD \psi \cD \bar\psi \cD A \,  e^{iS_{QCD}[\psi,\bar \psi, A]}
\lb{2.6e} \enq
and $S_{QCD}[\psi,\bar \psi, A]$ is the action, i.e.\ the integral
of the QCD Lagrangian (\rf{2.1}) over the space time continuum:
\beq
S_{QCD}[\psi,\bar \psi, A] =  \int d^3x d x_0 \LD_{QCD} \lb{2.7} \enq

If we consider the light quark sector of the theory and go to the
chiral limit where all quark masses are zero, the Lagrangian contains
no dimensionful quantity and is therefore formally scale invariant.
This scale is necessarily broken in order to regularize the Lagrangian and
the scale $\La_{QCD}$ is introduced which is the only scale in the chiral
limit. The dependence of the coupling constant 
$\al_s \equiv \frac{g_s^2}{4 \pi}$ on an arbitrary energy scale $\mu$ is defined
over the so called $\beta$ function :
\beq
\be(\al_s)=\mu \frac{\pa}{\pa \mu} \al_s(\mu).  \lb{4.3a}\enq
It can be calculated in perturbation theory and one obtains to lowest order:
\beq
\be(\al_s)= -\frac{\be_0}{2 \pi} \al_s^2 +O(\al_s^3) 
\qquad \mbox{ with } \be_0 = 11 - \frac{2}{3} n_f 
\lb{4.4} \enq
where $n_f$ is the number of relevant flavors.
The QCD scale $\La_{QCD}$ is now given as an integration constant of the
solution of equation (\rf{4.3a}) and again to lowest order we can write:

\beq \al_s(\mu) = \frac{4 \pi}{\beta_0 \ln(\mu^2/\La_{QCD}^2)} 
\qquad \mu \gg \La_{QCD} \lb{4.4a} \enq

This equation implies asymptotic freedom, since for large scales (high
momenta or correspondingly short distances) it vanishes logarithmically.

From this equation we get an intuitive argument that hadronic 
properties are inaccessible to calculation in  perturbation theory.  Let us try to calculate for
instance a scattering length $a_0$. 
It can get its dimension only by the the only dimensionful quantity, the
regularization scale $\mu$ and hence we would obtain: 
\beq
a_0 = f(\al_s) \mu^{-1} \enq
Off course the physical quantity
$a_0$ cannot depend on the arbitrary scale $\mu$ and we have:
\beq
\frac{d a_0}{d \mu} = f'(\al_s) \frac{1}{\mu}\frac{d\al_s}{d \mu} - \frac{1}{\mu^2}
f(\al_s) = 0 \enq
or equivalently:
\beq \frac{f'(\al_s)}{f(\al_s)} = \frac{1}{\be(\al_s)} \enq
which upon integration yields the expression:
\beq f(\al_s) = f(\al_0) \exp\left(\int_{\al_0}^{\al_s} \frac{d
\al'}{\be(\al')}\right) 
\enq
which cannot be expanded in a power series in $\al_s$ since the powers
of $\al_s$ are in the denominator of the exponent. 

\section{Methods and models in non-perturbative QCD}

Unfortunately we have for QCD - as for any other realistic quantum
field theory- only
one analytical method to treat the functional integral (\rf{2.6d}), namely perturbation theory.
 But as  mentioned
in the previous section, hadronic properties cannot be calculated with
perturbation theory, and therefore we have to rely here either on a
direct numerical simulation of the functional integral (\rf{2.6d}) or on models.
Models should respect the symmetries and explain the
most striking features of QCD : confinement and spontaneous
chiral symmetry  breaking. The former manifests itself by the fact,
that quarks and gluons have never been observed as free particles, the
second one, that we have the $\pi,\; K$ and $\et$ mesons as
pseudo-Goldstone particles in the hadronic sector.
For the numerical simulations one formulates the classical Lagrangian (\rf{2.1})
on a 4-dimensional hyper-cubic lattice in Euclidean space time,
preserving gauge invariance \ct{Wil74} and 
performs the high dimensional integrals numerically. With the
appearance of more and more powerful computers this approach gains more
and more importance and in some fields, as field theory at finite
temperatures it has replaced widely the experimental data. Indeed, lattice
people quite often refer to their results as measurements.

The oldest of the models -- it is older than QCD -- is the
non-relativistic potential model of quarks (see e.g. \cite{Gro91}). We can view the process of solving the 
the Schr\"odinger-equation of that model as one of the most powerful
tools for  performing  the
functional integration. Though potential models  give very satisfactory
results for 
hadron spectroscopy, the basic features, confinement and chiral
symmetry breaking have to put into the model and are not a consequence
of it. More sophisticated methods as attempts to solve the Schwinger Dyson 
equations or the Bethe Salpeter equation are much closer to the
original QCD, but I think it is fair to say that there is not yet a
real break-through to note.

Another important attempt  to solve the functional integral (\rf{2.6d}) is
based on 
the expansion of the action around classical solutions, the so called
instantons. 
This approach can explain very nicely chiral symmetry breaking, but not
confinement.

An approach based on modifying the QCD equations in order to make them
solvable but hopefully preserving the essential features is dual QCD.
It goes back to the proposals of t'Hooft\ct{tHo75} and Mandelstam
\ct{Man76} that the 
confinement mechanism in QCD is structurally similar to the formation
of Abrikosov-Nielsen-Olesen strings in a superconductor of the second
kind. In the 
latter case the condensation of Cooper-pairs implies that
magnetic fields cannot exist at all in superconductors of the first
kind (Meisner-effect) and only as flux tubes of magnetic fields in 
superconductors of the second 
kind, see Figure \rf{dual} a). 

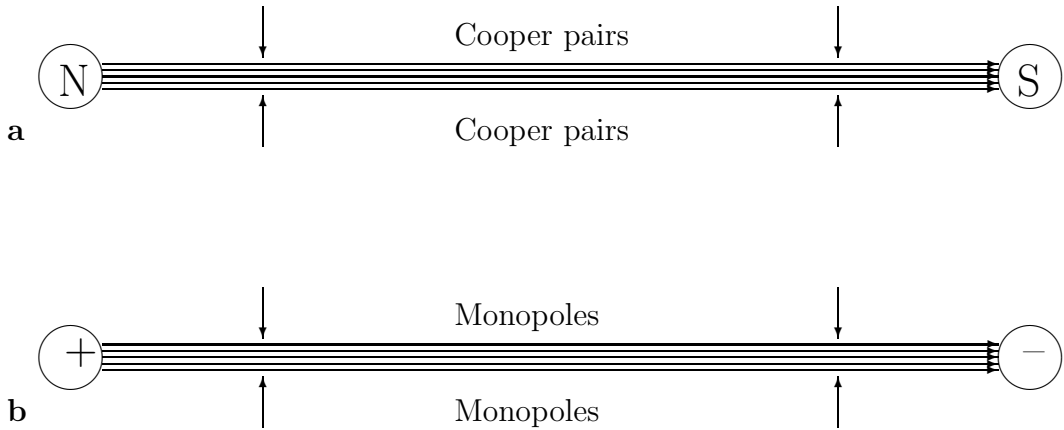
\begin{figure}[htb]
\setlength{\unitlength}{.085cm}

\begin{picture}(150,20)(-20,0)
\put(-10,0){\bf a}
\put(0,10){\circle{10}}
\put(150,10){\circle{10}}
\put(-2,7){\Large N}
\put(148,7){\Large S}
\put(5,12){\vector(1,0){140}}
\put(5,11){\vector(1,0){140}}
\put(5,10){\vector(1,0){140}}
\put(5,9){\vector(1,0){140}}
\put(5,8){\vector(1,0){140}}
\put(30,21){\vector(0,-1){8}}
\put(120,21){\vector(0,-1){8}}
\put(60,15){Cooper pairs}
\put(30,-1){\vector(0,1){8}}
\put(120,-1){\vector(0,1){8}}
\put(60,0){Cooper pairs}
\end{picture}

\vspace{2cm}

\begin{picture}(150,20)(-20,0)
\put(-10,0){\bf b}
\put(0,10){\circle{10}}
\put(150,10){\circle{10}}
\put(-1,9){\Large +}
\put(149,9){\Large --}
\put(5,12){\vector(1,0){140}}
\put(5,11){\vector(1,0){140}}
\put(5,10){\vector(1,0){140}}
\put(5,9){\vector(1,0){140}}
\put(5,8){\vector(1,0){140}}
\put(30,21){\vector(0,-1){8}}
\put(120,21){\vector(0,-1){8}}
\put(60,15){Monopoles}
\put(30,-1){\vector(0,1){8}}
\put(120,-1){\vector(0,1){8}}
\put(60,0){Monopoles}
\end{picture}

\caption{\lb{dual}The formation of a flux tube in a superconductor of second kind. 
a) electrodynamics, b)dual QCD}
\end{figure}

Magnetic  monopoles would thus be confined in an electric
superconductor of 
second kind, since the magnetic string between them would lead to a
linear increase of the field energy with growing distance. If QCD can
be approximated by a dual picture of the electrodynamics of a superconducter of
second kind, i.e. if color-magnetic monopoles condense and
color-electric flux tubes are formed, 
color charges are confined, see Figure \rf{dual} b).  
In the classical approximation the equations of motion imply the dual
potential $C^\mu$, the monopole condensate $\Phi$ and the field tensor
derived from a Dirac string of a quasi-static quark-antiquark pair
$G^S_{\al\be}$  which yields the coupling of the sources to the fields:
\beqa
\pa^\al(\pa_\al C_\be - \pa_\be C_\al ) &=& j_\be^{mon} - \pa^\al
G_{\al\be}^S \\
(\pa_\al - i g C_\al)^2 &=& -\frac{200 \la}{3}\Phi(|\Phi|^2 - B_0^2)
\nn
\enqa
with the monopole current:
\beq
j_\al^{mon} = - 3 i g[\Ph^*(\pa_\al - i g C_\al)\Ph 
- \Ph(\pa_\al + i g C_\al)\Ph^* 
\enq
Monopole condensation (in a maximal Abelian
gauge fixing) has been observed on the lattice, thus giving some
support to a kind of t'Hooft-Mandelstam mechanism, though confinement in
real QCD might be more subtle than in the dual picture \ct{DiG97}. There has been a recent revival of interest in dual QCD since
in supersymmetric QCD \ct{SeiW94,Sei95} the duality transformation can
be performed and the long distance behaviour of supersymmetric QCD  is
described by a weekly coupled dual gauge theory.  I cannot go into
details of dual 
QCD and refer to the papers of  Ball, Baker, Zachariasen
and coworkers (see \ct{BBBPZ96} and the literature quoted there) who
have elaborated the model in great detail . 

The last model I want to present shortly is the model of the stochastic
vacuum (MSV) proposed by Yuri Simonov and myself several years ago
\ct{Dos87,DosS88}, for reviews see \ct{Sim96,Dos97}. It
starts from the assumption that the complicated structure of the
colordynamic interaction at long distances  can be approximated by a
simple stochastic 
process, in the most restricted and phenomenologically most useful form
by a Gaussian process, which is determined by the gauge invariant
correlator of two color- field strength tensors $F^C_{\mu\nu}$ at
different space time points. 

\begin{eqnarray}\lb{M.21}
&&<F^F_{\mu\nu}(x,y)
F^G_{\kappa\lambda}(0,y)>=\frac{\delta^{FG}}{12(N^2_c-1)}<FF>
\cdot\left\{(\delta_{\mu\kappa}\delta_{\nu\lambda}
-\delta_{\mu\lambda}\delta_{\nu \kappa}) D(x^2)\cdot \kappa\right.\\
&& \quad+\left.\left[\frac{1}{2}\frac{\partial}{\partial
x_\mu}(x_\kappa\delta_{\nu\lambda}-x_\lambda \delta_{\nu
\kappa})+ 
\frac{1}{2}\frac{\partial}{\partial
x_\nu}(x_\lambda\delta_{\mu\kappa}-x_\kappa
\delta_{\mu\lambda})\right]D_1(x^2)(1-\kappa)\right\}.\nn
\end{eqnarray}

Here we have switched to an Euclidean space time with $x_4 = i \,x_0$
and $y$ is a reference point to which the color content has been
transported. A nice feature of the model is that it
leads  to linear confinement for non-Abelian theories, but
not to confinement for Abelian gauge theories like QED, unless there is
monopole condensation. It also reproduces correctly the spin structure
of the confining potential between heavy quarks. There are new
developments towards its application to light quarks and an
understanding of chiral symmetry breaking \ct{Sim97}. Confinement in
this model 
is the result of the special tensor structure in front of $D(x^2)$ in
the correlator  (\rf{M.21}). This structure
 is forbidden in an Abelian theory by the
homogeneous Maxwell equations, but can be present in a non-Abelian gauge theory
like QCD. It had been shown  by Di Giacomo and coworkers \cite{DiGP92,DiGMP97}
on the
lattice that this confining tensor structure is indeed present and even
dominant in QCD.  

\section{Formation of a color-electric flux tube in QCD }

String formation between static quarks has been observed in lattice
calculations (for SU(2) 
as gauge group) \ct{HayP93,BSS95,DGS94}. In Figure \rf{lastri1}
taken from reference \cite{BSS95} the 
the action density (i.e. the difference between the squared electric
and magnetic field strength) for a static quark antiquark pair at
different separations $R$ is displayed. One clearly sees the
contribution of the color-electric Coulomb field  around the quark and
anti-quark, but also the connecting string is clearly visible. 
\epsfxsize16cm
\begin{figure}[htb]
\leavevmode
\centering
\epsffile{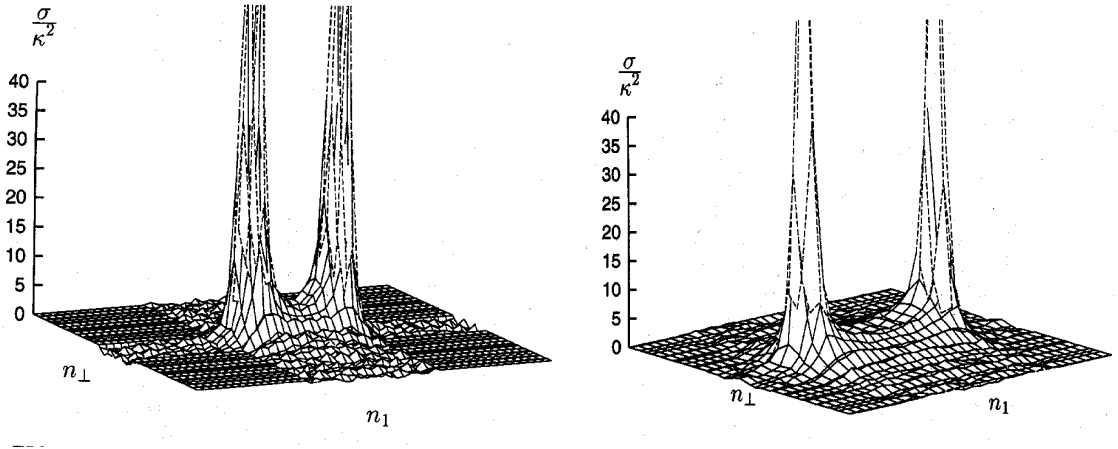}
\caption{\lb{lastri1} Action density (in units of the string tension)
of a quark antiquark pair at distances of ca 0.7 and 1.35 fm. The figure
is taken from \protect \cite{BSS95}.} 
\end{figure}

The formation of a colorelectric flux tube is
a direct consequence in dual QCD see Figure \rf{dual}. Also the model
of the stochastic vacuum  
leads to  the formation of a colorelectric flux tube \ct{DGS94,RueD95}.
In Figure \rf{msvstri1} taken from \cite{RueD95} the squared
color-electric field strength as calculated in the model of the
stochastic vacuum is displayed. The ``Coulomb peaks'' around the
positions of the quark and anti-quark are due to the Coulomb
contribution to correlator $D_1$ of equation (\rf{M.21}), the
connecting string however is entirely due to the typically
non-perturbative and non-Abelian structure $D$. 

\epsfxsize15cm
\begin{figure}[htb]
\leavevmode
\centering
\epsffile{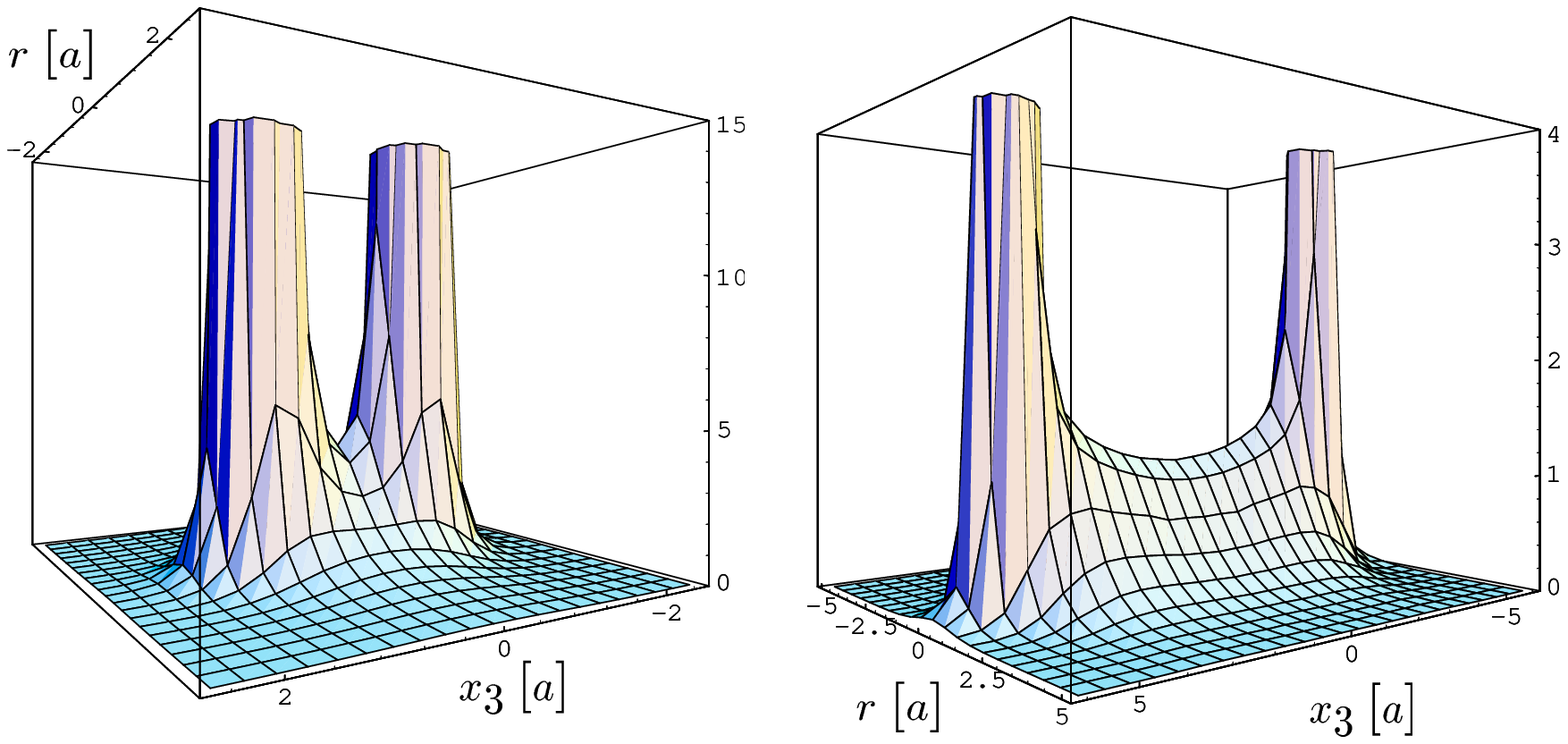}

\caption{\lb{msvstri1} The squared color-electric field of a static quark-antiquark pair at distances ca. 0.7 and  3 fm  as calculated in the model of the stochastic vacuum, figure taken from \protect \cite{RueD95}.}
\end{figure}
The excitations of the string, or at least its long distance
consequences have been observed in lattice calculations; the two
models treat, at least up to now, only a static string. The full width
of the string is similar in all cases, about one fermi ( a bit less in
the SU(2) case). The width of the string is nearly independent of the
distance between the quark and antiquark, the same holds for the energy
density thus leading to a linearly rising potential energy between the
static sources, see Figure \rf{msvstri2} for the situation in the model of the stochastic vacuum. 

\epsfxsize15cm
\begin{figure}[htb]
\leavevmode
\centering
\epsffile{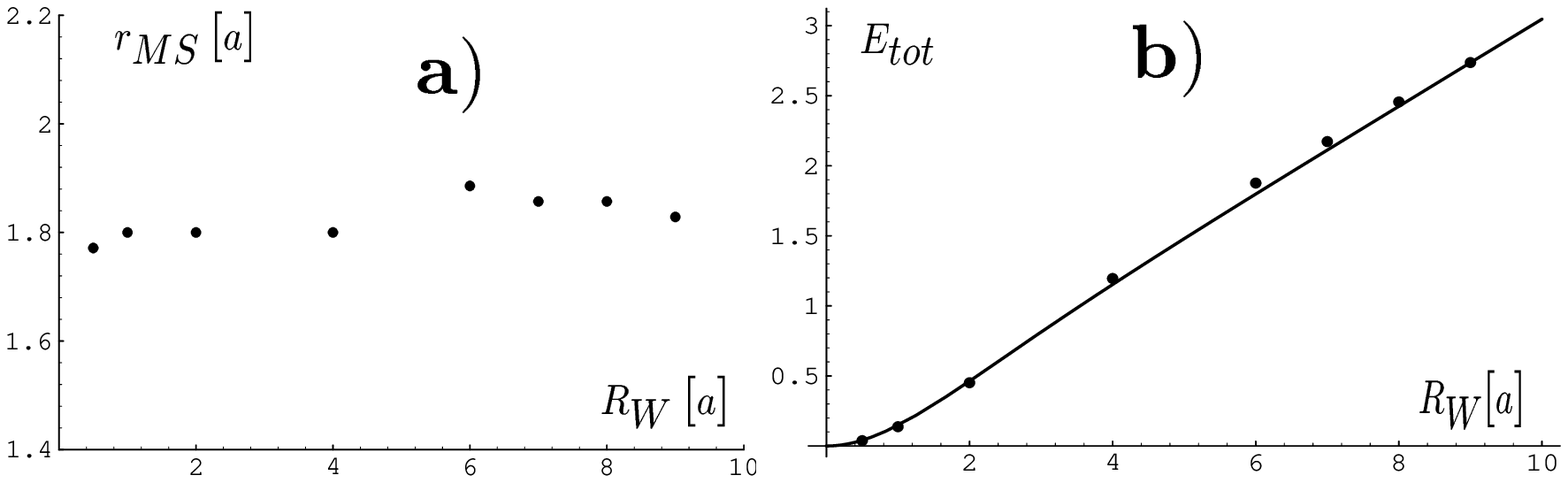}
\caption{a) The width of the string as a function of the distance of the static quark-antiquark pair, the unit $a$ is about 0.3 fm, from \protect \ct{RueD95},
b)The potential energy of a static quark-antiquark pair  calculated by the area law of the Wegner-Wilson loop (solid line) and calculated as color-electric field energy (dots) from \protect \ct{RueD95}.\label{msvstri2}}
\end{figure}
The consequences of string formation for spin and
velocity dependent forces between heavy quarks  have been studied
intensively by Brambilla and Vairo \ct{BraV97} leading to very similar
results for the MSV and dual QCD.  

If we regard interactions between more than two particles, string
formation tells us that we cannot expect that many body forces can be obtained by adding just the two body forces. In general a string is is saturated if it has
found its two end points. This implies that in contradistinction to
electrodynamics we cannot count with van
der Waals forces between hadrons which fall off like an inverse power,
at least if the distances between the 
hadrons are comparable or larger than the string width. This is off course in agreement with the general field theoretical result that forces between hadrons fall off exponentially due to the finite masses of the exchanged hadrons.

 But before I
come to hadron-hadron interactions let me shortly discuss the nucleon
in non-perturbative QCD and exotics.

\subsection{Nucleons and exotics}
Mathematically a string between colored objects can be expressed by a
generalization of the Schwinger string of electrodynamics:
\beq \lb{string}
S(x,y) := P \exp\left(- \frac{i g}{2}  \int_{C(x \to y)} (A^F_\mu \la_F) dz^\mu 
\right)
\enq
where $C(x \to y)$ is a curve connecting the points $x$ and $y$, $\la_F$
are for SU(3) the Gell-Mann matrices and P denotes path ordering, that
is the prescription how to treat the integral over non-commuting
quantities in the exponential. $S(x,y)$ is an element of SU(3), and it has
the important property to transport the information on color from
point $x$ to point $y$. Due to local gauge invariance it makes no sense
to speak e.g. of color neutrality of a nonlocal object, like 
$\sum_C  \bar \psi^C(x)\psi^C(y)$, unless we insert the Schwinger string
$S(x,y)$ between it (or use a gauge, in which $S(x,y)$ is unity).

A special feature of SU(N) is that one can couple $N$ objects
transforming under the fundamental transformation to a singlet, in case
of SU(3) thus three quarks. This is a trivial consequence of the fact
that the determinant of an {\bf S}U(N) matrix is by definition one, i.e.
\beq \lb{eps}
 \epsilon_{ikl} U_{ir} U_{ks} U_{lt} = \ep_{rst} 
\enq
if $U_{ab}$ is a special unitary 3$\times$ 3 matrix. 
This allows to form a color neutral 3 quark state and was an
essential ingredient of the quark respective aces- model of Gell-Mann
and Zweig. In the string picture the baryon of three quarks is formed
just by joining three Schwinger strings according to \rf{eps} at the
point $y$:
\beq \lb{nucleon}
\psi^C(x) \psi^D(u) \psi^E(w) S_{C F}(x,y) S_{D F}(u,y) S_{E F}(u,y)
\enq
If we assume that the strings carry energy proportional to their length and that the width can be neglected ( a rather unrealistic assumption)
one can derive the genuine 3-quark potential inside a nucleon, see Figure 
\rf{nucl}:
\epsfxsize9cm
\begin{figure}[htb]
\leavevmode
\centering
\epsffile{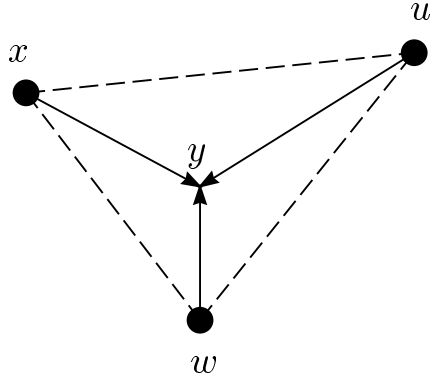}
\vspace{-2cm}
\caption{\lb{nucl}The genuine 3 body potential of the constituent quarks at positions $u\,w\,x$ in the baryon as addition of the three string lengths emerging from the point $y$. The dashed lines indicate the approximation by addition of the three lines of the circumference, see equation \protect \rf{barax}}
\end{figure}
\beq \lb{tori}
V_N(\vec x,\vec u,\vec w) = \min_y(|\vec x - \vec y| +|\vec u - \vec y|+
|\vec w - \vec y|)
\enq
The point $y$ for which the rhs of \rf{tori} is minimal has been found by
Toricelli. If the triangle spanned by the points $\vec x, \vec u,
\vec w$ has no angle larger than $2 \pi/3$, then $y$ is an inner point
of it and its connecting lines to the corners form a star with angles
$2 \pi/3$. Otherwise, if one of the angles of the triangle is greater
than $2 \pi/3$ $y$ is the corner point with this angle.

Simple geometrical considerations give
\beq \lb{barax}
\min_y(|\vec x - \vec y| +|\vec u - \vec y|+
|\vec w - \vec y|) = f(\vec x, \vec u, \vec w) \{|\vec x - \vec u|
+|\vec u - \vec w|+ |\vec w - \vec x|\}
\enq
with 
\beq
1/2 \leq f(\vec x, \vec u, \vec w) \leq 1/\sqrt{3} \;(\approx 0.58)
\enq
Thus if the string tension is $\sigma$ for the quark antiquark pair,
the 3-quark potential in the nucleon (\rf{tori}) can be quite well
approximated (to 
better than 10 \%) by the additive effective quark-quark potential 
\beq \lb{eff}
V_N(\vec x,\vec u,\vec w) \approx  V_{eff}(\vec x,\vec u)+
V_{eff}(\vec u,\vec w)+
V_{eff}(\vec w,\vec x), \qquad
V_{eff}(\vec x,\vec u) = 0.54 \si |\vec x - \vec u|
\enq
These considerations were first put forward by V. M\"uller and
myself\ct{DosM76} 
inspired from strong coupling expansion of lattice QCD \ct{Wil74}.
Calculations of baryonic spectra based on the effective potential (\rf{eff})
turned out to be very reasonable see \ct{Gro91}, much more elaborate
calculations with very good results were also performed on the basis of
similar considerations by Fabre de la Ripelle and Simonov \ct{FabS91}. 

In the string picture we may also construct more elaborate states than
mesons and baryons, as $(qq)-(\bar q \bar q)$ or 6 quark states as
indicated in Figure \rf{exo}.
\epsfxsize14cm
\begin{figure}[htb]
\leavevmode
\centering
\epsffile{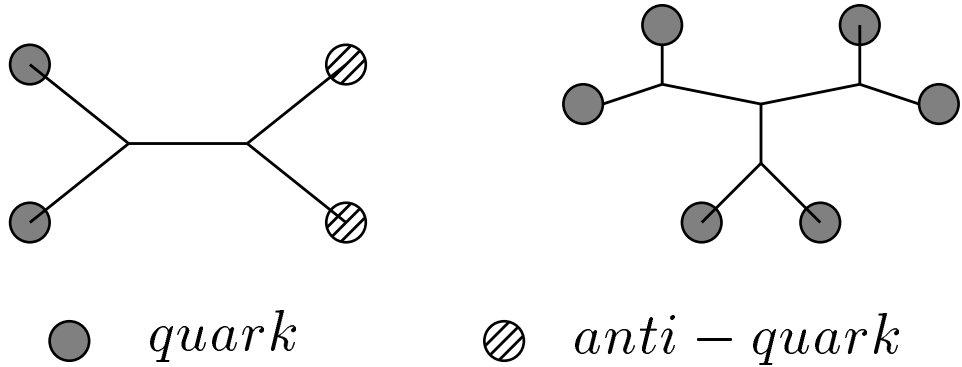}
\caption{\lb{exo}Exotic states constructed in the string picture}
\end{figure}

Spectroscopy has up to now given no compelling reason
for the existence of these states. At the moment all existing
resonances seem to be compatible to be quarkonia, glue-balls or
superpositions thereof \ct{Nar97}.

\section{High energy scattering and few-body problems}
\subsection{Quark-additivity versus string-string interaction}
In baryon spectroscopy we have 3 valence quarks involved, in meson
baryon scattering at least 5 of them. For low energy scattering one
cannot dispose of meson exchange and one comes therefore in problems
which are extremely intricate from the QCD point of view. For high
energy scattering one expects that the leading contribution to elastic
scattering is determined by gluon exchange \ct{Low75,Nus75}, 
so there is some hope that with non-perturbative methods one can treat
the problem. From the few-body point of view, there is an interesting question:
Is the hadron hadron scattering determined by the valence quarks, or
does the string also participate at the interaction. In an Abelian
Model Landshoff and Nachtmann \ct{LanN87} found that only the quarks
participate and if the distance of
the quarks inside a hadron is large as compared to the correlation
length of the gluon field strength correlator introduced above, the
total hadron cross section is a superposition of the quark cross
sections, justifying in such a way the quark additivity found in the
early days of the quark model \ct{LevF65,LipS66}. If we have
additivity of the quark cross sections the pion-nucleon cross section
is 6 times , 
the nucleon- nucleon cross section 9 times the quark-quark cross section .
The so predicted ratio 2/3 for $\pi-N$ to $N-N$ scattering is indeed
observed for the total cross sections at high energies. 

In the model of the stochastic vacuum, where the non-Abelian nature of
QCD plays a crucial role, we can also treat high energy scattering \ct{DosFK96},
using more or less the same formalism which was applied to find the
confining potential or the color-electric flux tube . The
essential ingredients are expectation values of two Wegner-Wilson loops with
lightlike sides, see Figure \rf{2loop}. 
\epsfxsize14cm
\begin{figure}[htb]
\leavevmode
\centering
\epsffile{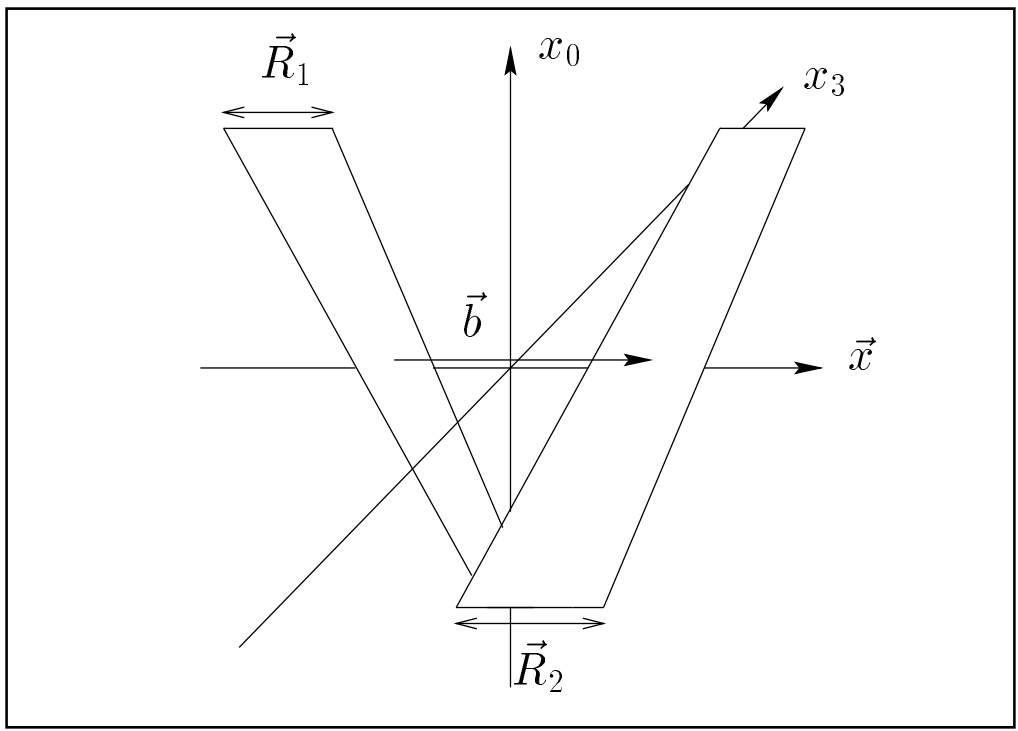}
\caption{\lb{2loop} The building block of hadron hadron scattering in
the model of the stochastic vacuum. The lightlike sides of the two
Wegner-Wilson loops represent the paths of the quark and antiquark
forming a color dipole. The expectation value of these two loops can be
calculated in the framework of the model of the stochastic vacuum.
Hadron-hadron scattering is obtained from dipole-dipole scattering by
smearing over the transversal extension of the loops with transversal
hadronic wave functions} 
\end{figure}

A crucial result of the model is that the same
mechanism which leads to string formation and confinement also leads to
a string string interaction, thus invalidating quark-additivity.
Nevertheless also this model yields the correct ratio of $\pi-N$ to
$N-N$ scattering without any new free parameter, since it is determined
by the known (electromagnetic) radii of the pion and the nucleon. It
furthermore predicts the ratio of $K-N$ to $\pi-N$ scattering correctly 
also from the elctromagnetic radii, without introducing a different
cross section for scattering of strange and non-strange quarks as has to
be done in the quark additivity scheme and which is strange to the
flavor independence of the quark-gluon interaction, see Table \rf{tab}.
\begin{table}
\centering
\begin{tabular}{|c|c|c|}  \hline
Ratio & Model & Exp.\\  \hline
$\sigma_{p\pi}/\sigma_{pp}$ & $0.66\pm 0.02$ &  0.63\\ 
$\sigma_{p K }/\sigma_{p\pi }$ & $0.82\pm 0.08$ & 0.87\\
\hline
\end{tabular}
\caption{\lb{tab}Predictions of the model of the stochastic vacuum for total
$\pi$-nucleon and $K$-nucleon cross sections, from \protect \ct{DosFK96}} 
\end{table}
An  essential feature of the string string interaction is the increase
of the cross 
section with increasing hadron size. In electro-production of
vector-mesons the virtuality $Q^2$ of the photon is the handle for
constructing quark-antiquark states with different sizes 
$R_{q\bar q} \propto \frac{1}{Q}$. Nemchik {\em et al.} \ct{Nem96}
have disentangled the cross section of a quark-antiquark state
(color-dipole) of a certain size $R_D$ from data of electro-production
of vector mesons from protons.  Their results are shown in Figure \rf{nofit}.
The dashed line is their phenomenological fit to the dipole-proton
cross section, the solid line is our prediction from the model of the stochastic vacuum, where no high-energy
data have been used \ct{RueD97}. The data show no sign of saturation at dipole
sizes small as compared to the hadron radii, as would be necessary for
quark additivity to hold and our prediction fits the semi-experimental
results within the expected accuracy.
\epsfxsize10cm
\epsfysize8cm
\begin{figure}[ht]
\leavevmode
\centering
\epsffile{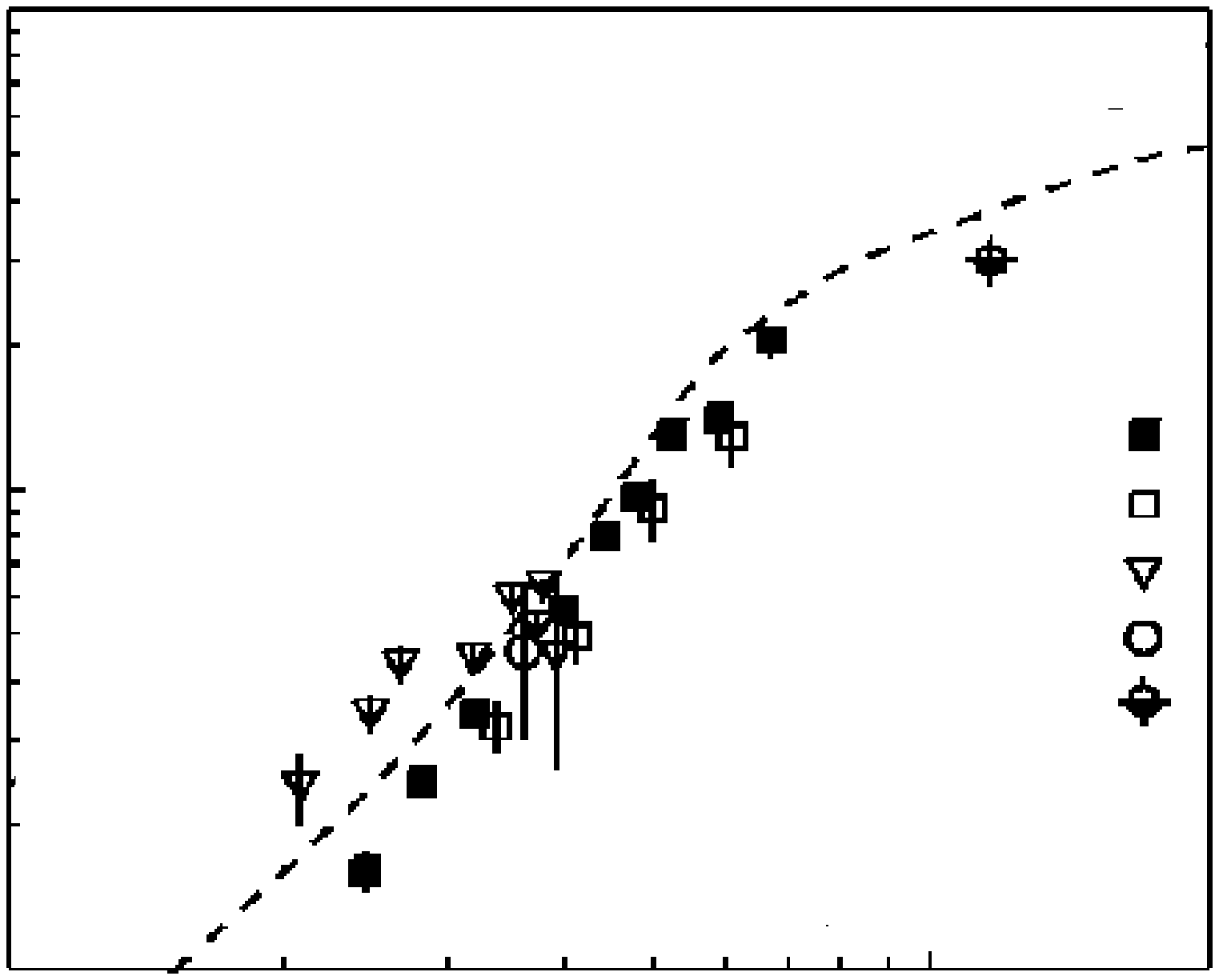}
\begin{picture}(0,0)
\put(-280,5){\epsfysize7.65cm\epsfxsize9.63cm\epsffile{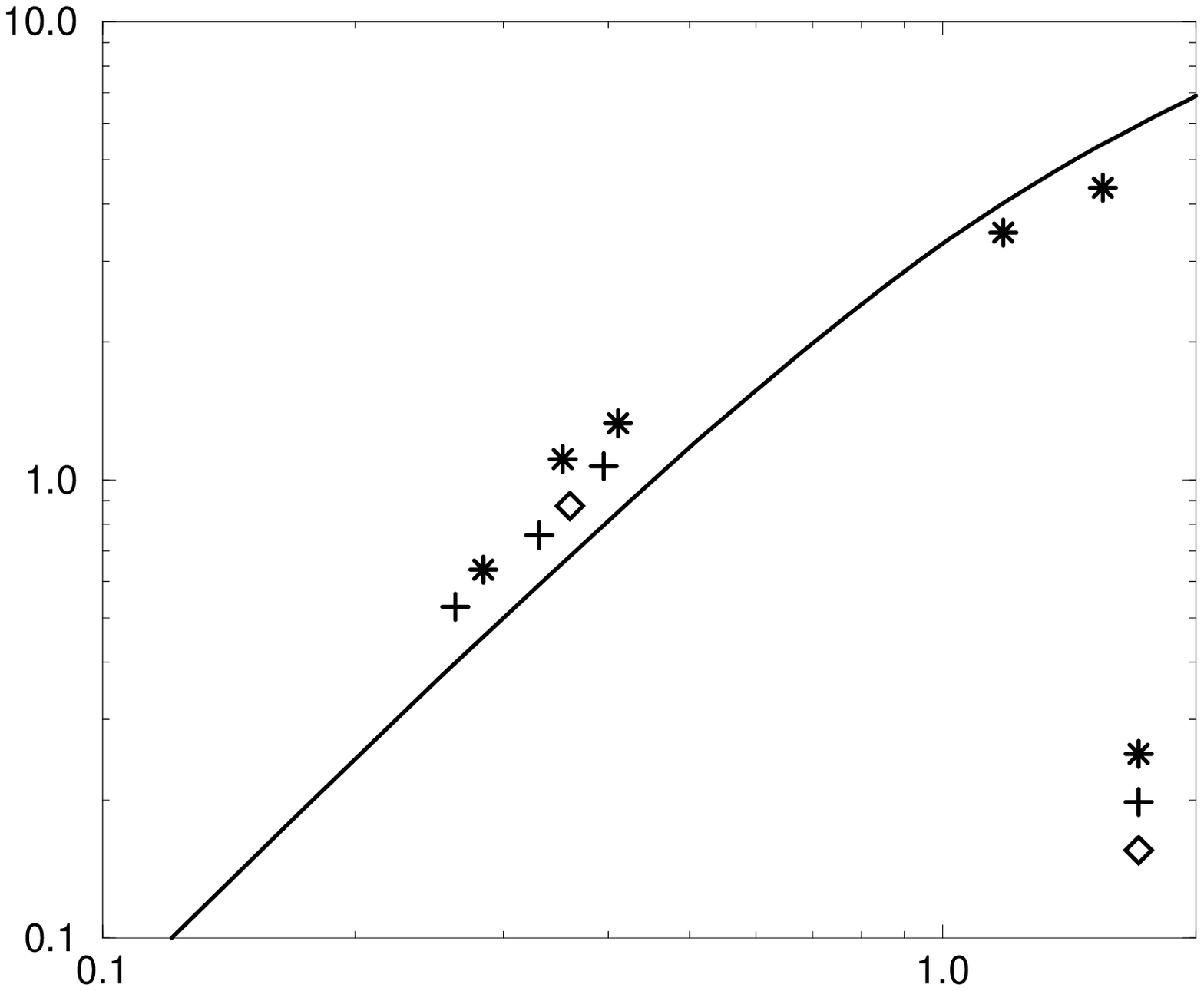}}
\put(-1.5,0){$R_{{\rm D}}[{\rm fm}]$}
\put(-330,205){$\si^{\rm tot}[{\rm fm}^2]$}
\put(-95,125){$\rh^0$-NMC}
\put(-95,110){$\Ph^0$-NMC}
\put(-95,95){$J/\Psi$-EMC}
\put(-95,80){$J/\Ps$-E687}
\put(-95,65){$\Ph^0$-FNAL}
\put(-95,50){$\Ph^0,\,\rh^0$-ZEUS}
\put(-95,35){$\rh^0$-H1}
\put(-95,20){$J/\Ps$-HERA}
\end{picture}
\begin{minipage}{16cm}
\caption{ \lb{nofit} Comparison of our result for the total cross section for dipole (extension $R_{{\rm D}}$) proton scattering with values extracted from cross sections of lepto-production of vector-mesons by the method of Nemchik et al.~\protect\cite{Nem96}. The solid line is our result without any fitting of the parameters to high-energy data. The dashed line is the ansatz of Nemchik et al.~for the total cross section,\protect \ct{RueD97}.}

\end{minipage}
\end{figure}

\subsection{Nucleon Structure and the C=P=-1 Exchange}
If the leading contribution to high energy scattering is due to gluon
exchange, there is no compelling reason, why the exchange with vacuum
quantum numbers, i.e. the one with positive parity and positive charge
parity (C=P=+1) should be overwhelmingly dominant. Indeed if the
leading contribution to the C=P=+1 exchange is a ``non-perturbative two
gluon state'' a ``non-perturbative three gluon state'' should not be
terribly suppressed and such a state can also have the quantum numbers
C=P = {\bf -} 1. The exchange of such an object has been considered first by Lukaszuk and
Nicolescu \ct{LukN73} and been called odderon. Such a state would
contribute differently to proton-proton and proton-antiproton
scattering and lead to a difference in the ratio of the real to the
imaginary part of the respective scattering amplitudes. However such a
difference has not been observed, experimentally the difference 
\beq \lb{deltarho}
\De\rh(s) =\frac{\mbox{Re}[T^{\bar p p}(s,t=0)]}
{\mbox{Im}[T^{\bar p p}(s,t=0)]}
- \frac{\mbox{Re}[T^{ p p}(s,t=0)]}{\mbox{Im}[T^{ p p}(s,t=0)]}
\enq
is at $\sqrt{s}$ = 546 GeV \ct{UA493} smaller than approximately 0.02
and well compatible with zero. This indicates a strong suppression of the
odderon. Such a behavior is hardly compatible with an additive quark
model, and indeed Donnachie and Landshoff \ct{DonL91} have found in an
Abelian model for the pomeron \ct{LanN87} a value of $\De\rh$  of about
0.5. The model of the stochastic vacuum applied to high energy
scattering would yield a similar value if the nucleon is assumed to be a
fully symmetric 3 quark structure. There are however several
indications that two quarks in the nucleon cluster to a diquark . In
such a case this diquark behaves as far as color is concerned like an
antiquark. The odderon may couple to such a state, but if we rotate the
quark-diquark state by $\pi$ we obtain, due to the negative parity of
the odderon, a contribution of opposite sign which cancels the
unrotated one. The effect has been discussed quantitatively in reference
\ct{RueD96} and the following results have been found. If the three
quarks sit at the corners of an triangle and we decrease the baseline $r_\perp$
the C=P=-1 contribution decreases dramatically leading to a $\De\rh
\leq .02$   for a diquark radius of ca. 0.3 fm, see Figure \rf{tria}. Thus even a moderatly
extended diquark yields the desired odderon suppression.
\epsfxsize9cm
\begin{figure}[htb]
\leavevmode
\centering
\epsffile{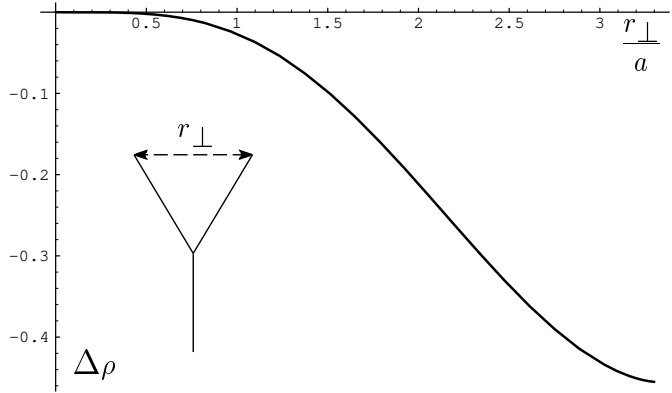}
\caption{\lb{tria}The difference $\Delta \rho$, see equ. \protect (\rf{deltarho})
as a function of the ``diquark radius'' \protect $r_\perp$, \protect \ct{RueD96}.}
\end{figure}
This mechanism can be tested further at Hera. If one considers the
photo or electro-production of pseudo-scalars and pseudo-vectors the
odderon will couple to the photon-meson transition. If the nucleon has
the diquark structure as discussed above, it cannot couple to a proton
directly, but to a proton-baryon$(J^P=\frac{1}{2}^{-1})$ transition vertex, see Figure \rf{odd}.
\epsfxsize9cm
\begin{figure}[htb]
\leavevmode
\centering
\epsffile{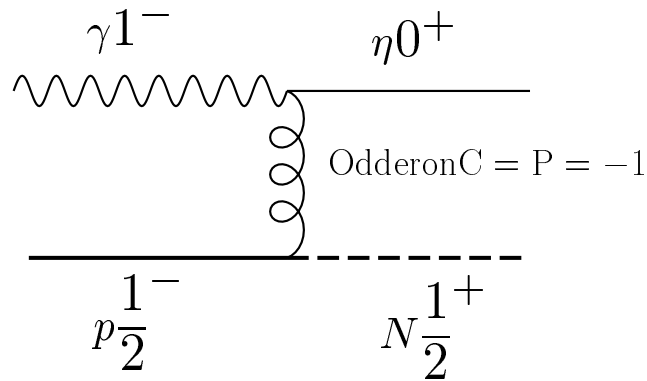}
\caption{\lb{odd} Odderon exchange contribution to diffractive photo- or electro-production of odd parity states off nucleons}
\end{figure}
A good candidate for the isolation of the odderon would thus be a
reaction like:
$$ \ga^{(*)} + p -> \et + N(1535)(\frac{1}{2}^-)$$ 
in the diffractive
region. The signatures would be the decay gammas of the $\et$ and the
decay neutron of the N(1535). 

\section{Acknowledgments}
It is a pleasure to thank the organizers of the conference for creating a pleasant and stimulating atmosphere. I am grateful to N. Brambilla, A. Di Giacomo, D. Gromes, E. Meggiolaro, O. Nachtmann, M. Rueter, and A. Vairo for informative discussions.

\end{document}